\documentclass[twoside]{ilcws07}
\usepackage[latin1]{inputenc}
\usepackage[dvips]{graphicx,epsfig,color}
\usepackage{wrapfig,rotating}
\usepackage{amssymb,amsmath,array}
\usepackage{cite} 
\def\lsim{\raise0.3ex\hbox{$\;<$\kern-0.75em\raise-1.1ex\hbox{$\sim\;$}}}
\begin{document}
\title{
Polarization aspects \\in radiative neutralino production}
\author{Herbert K. Dreiner, Olaf Kittel\thanks{Speaker}, and Ulrich Langenfeld
\vspace{.3cm}\\
Physikalisches Institut der Universit\"at Bonn \\
Nu\ss allee 12, 53115 Bonn - Germany
}

\maketitle

\begin{abstract}

We study the impact of beam polarization on radiative neutralino 
production $e^+e^- \to \tilde\chi^0_1\tilde\chi^0_1\gamma$
at the International Linear Collider. We show that
longitudinal polarized beams significantly enhance the signal 
and simultaneously
reduce the Standard Model background from
radiative neutrino production $e^+e^- \to \nu \bar\nu \gamma$.
We point out that 
the radiative production of neutralinos could be
the only accessible SUSY particles, if neutralinos, 
charginos, sleptons, as well as squarks and gluinos are too heavy
to be pair-produced in the first stage of the ILC
at $\sqrt{s} =500$~GeV.
\begin{picture}(10,10)
\put(45,260){BONN-TH-2007-04}
\end{picture}
\end{abstract}

\section{Introduction}

The Minimal Supersymmetric Standard Model (MSSM) is a promising
extension of the Standard Model of particle physics (SM)~\cite{Haber:1984rc}.
At the International Linear Collider (ILC)~\cite{ilc},
the masses, decay widths, couplings,
and spins of the new SUSY particles can be measured with high
precision~\cite{Aguilar-Saavedra:2005pw}. 
In particular, the lightest states 
like pairs of neutralinos,
charginos, and sleptons, 
can be studied in the initial stage of the ILC,
with a center-of-mass energy $\sqrt s = 500$~GeV, and a 
luminosity of ${\mathcal L}=500$~fb$^{-1}$.
The lightest SUSY state is a pair of 
radiatively produced neutralinos~\cite{Datta:1994ac,Ambrosanio:1995it,Datta:1996ur,
Choi:1999bs,Baer:2001ia,Datta:2002jh,Dreiner:2006sb,Dreiner:2007vm}
\begin{equation}
e^+ + e^-\to\tilde\chi_1^0 + \tilde\chi_1^0 + \gamma. 
\label{neut-pairs}
\end{equation}
The signal is a single high energetic photon, radiated off the incoming 
beams or off the exchanged selectrons,
and missing energy, carried by the 
neutralinos~\cite{Wilson:2001me,Birkedal:2004xn,Bartels:2007ag}.

\section{Signal and background}

The main Standard Model background is photons from
radiatively  produced 
neutrinos $e^+e^- \to \nu \bar\nu \gamma$~\cite{Birkedal:2004xn,Bartels:2007ag}.
In order to quantify whether
an excess of signal photons from radiative
neutralino production, $N_{\mathrm{S}}=\sigma {\mathcal L}$, 
can be observed 
over the SM background photons, $N_{\rm B}=\sigma_{\rm B}{\mathcal L}$,
we define the theoretical significance $S$,
and the signal to background ratio  $r$~\cite{Dreiner:2006sb}
\begin{equation}
S  =  \frac{N_{\rm S}}{\sqrt{N_{\rm S} + N_{\rm B}}}=
\frac{\sigma}{\sqrt{\sigma + \sigma_{\rm B}}} \sqrt{\mathcal L}, \qquad 
r =  \frac{N_{\rm S}}{N_{\rm B}}=\frac{\sigma}{\sigma_{\mathrm{B}}}.
\end{equation}
For example, a theoretical significance of $S = 1$ implies that the signal
can be measured at the statistical 68\% confidence level.  
If the experimental error of the background cross section is 1\%,
the signal to background ratio must be larger than 1\%.
A detection of the signal requires at least
\begin{equation}
S > 1 \quad \mathrm{and}\quad r>1\%.
\end{equation}  

\section{Cuts on photon angle and energy}

For the tree-level calculation of the cross sections $\sigma$ 
for signal and background, we use the
formulas for the amplitudes squared as given in Ref.~\cite{Dreiner:2006sb}.
To regularize the infrared and collinear divergences, we apply cuts on
the photon scattering angle $\theta_\gamma$
and energy $E_\gamma$~\cite{Dreiner:2006sb}
\begin{equation}
 |\cos\theta_\gamma| \le 0.99,\qquad 
0.02 \le x\le \;1-\frac{m_{\chi_1^0}^2}{E_{\rm beam}^2},
\qquad
x = \frac{E_\gamma}{E_{\rm beam}}, 
\label{cuts}
\end{equation}
The upper cut on the
photon energy $x^{\mathrm{max}}=1-m_{\chi_1^0}^2/E_{\rm beam}^2$ is
the kinematical limit of radiative neutralino production. This cut also
reduces much of the on-shell $Z$ boson contribution to the background
from radiative neutrino production~\cite{Dreiner:2006sb}.
Note that the ratios $r$ and $S$ do not depend very sensitively on the
choice of the cuts $|\cos\theta_\gamma| \le 0.99$ and $0.02 \le x$,
since signal and background 
have very similar distributions in energy $E_\gamma$ and angle $\theta_\gamma$.

\section{Numerical Results}

We present numerical results of the signal and background
cross sections with emphasis on their dependence on the  
beam polarization~\cite{Moortgat-Pick:2005cw}, and on the higgsino and gaugino mass 
parameters $\mu$ and $M_2$, respectively.
Finally we discuss a mSUGRA scenario, where
radiative production of neutralinos is
the only accessible SUSY state 
at $\sqrt{s} =500$~GeV.

\subsection{Beam polarization dependence}

\begin{figure}[t]
\begin{picture}(200,100)
\put(-100,-260){\includegraphics{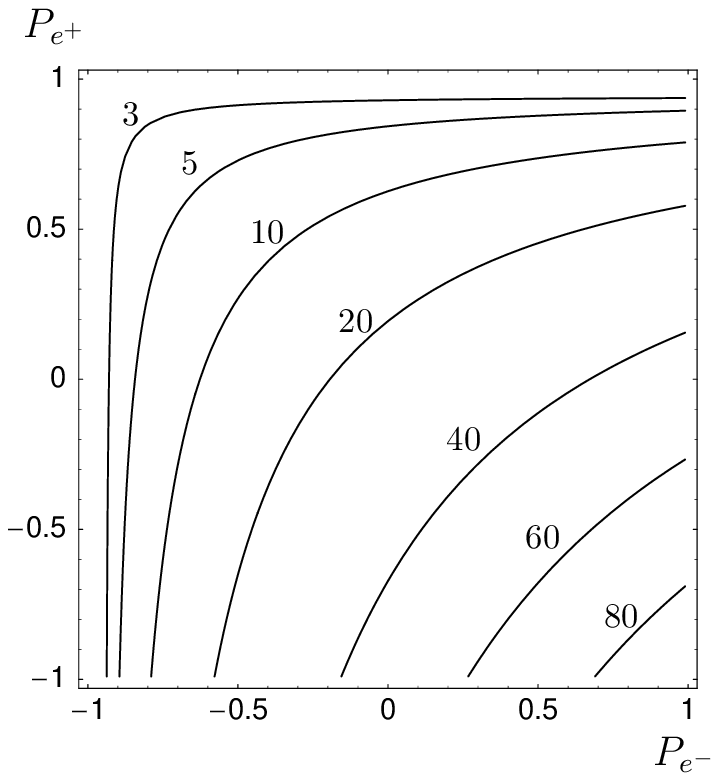}}
\put(40,-260){\includegraphics{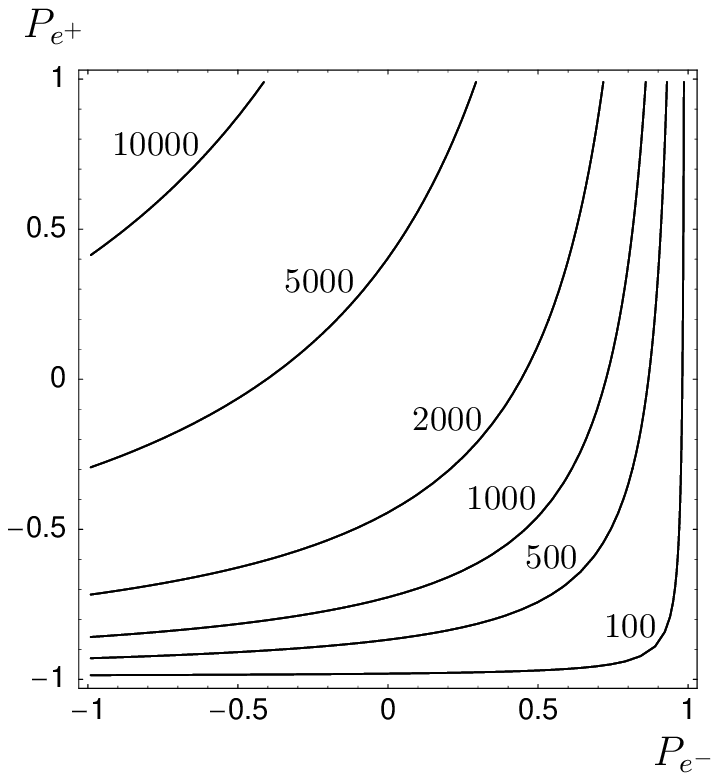}}
\put(180,-260){\includegraphics{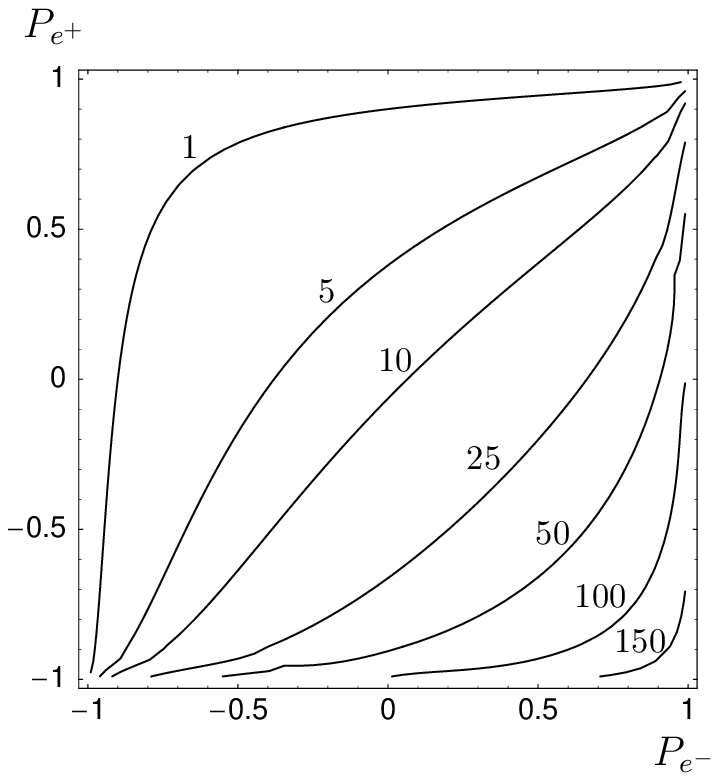}}
\put(10,125){$\sigma(e^+e^-\to\tilde\chi_1^0\tilde\chi_1^0\gamma)$ in $\rm{fb}$}
\put(160,125){$\sigma_{\rm B}(e^+e^-\to\nu\bar\nu\gamma)$ in $\rm{fb}$}
\put(310,125){Significance $S$}
\end{picture}
\caption{\small 
        Dependence of the cross sections and the significance
        on the beam polarizations $P_{e^-}$, $P_{e^+}$ with
        $\sqrt s = 500$~GeV, ${\mathcal L}=500$~fb$^{-1}$
        for SPS1a with
        $\mu=352$~GeV, $M_2=193$~GeV, $\tan \beta=10$, 
        $m_0=100$~GeV, $m_{\tilde\chi^0_1}=94$~GeV, and
        $m_{\tilde\ell_{R(L)}}=143(204)$~GeV~\cite{Dreiner:2006sb}.
}
\label{fig:beampol}
\end{figure}

In Fig.~\ref{fig:beampol}, we show the beam polarization dependence
of the signal $\sigma(e^+e^-\to\tilde\chi^0_1\tilde\chi^0_1\gamma)$
and background cross sections $\sigma_{\rm B}(e^+e^-\to\nu\bar\nu\gamma)$.
In  the SPS~1a scenario~\cite{Allanach:2002nj},
the neutralino is mostly bino, such that
radiative neutralino production dominantly proceeds
via right selectron $\tilde e_R$ exchange.
The background, radiative neutrino production,
mainly proceeds via $W$ boson exchange. Thus
positive electron beam polarization $P_{e^-}$ and negative positron
beam polarization $P_{e^+}$ enhance the signal cross section and
reduce the background at the same time, such that the significance
is greatly enhanced, see Fig.~\ref{fig:beampol}.
For beam polarizations of $(P_{e^-}, P_{e^+})=(0.8,-0.3)[(0.8,-0.6)]$,
the signal cross section is
$\sigma(e^+e^-\to\tilde\chi^0_1\tilde\chi^0_1\gamma)=56[70]$~fb,
the background is $\sigma_{\rm B}(e^+e^-\to\nu\bar\nu\gamma)=540[330]$~fb,
such that the significance is $S=50[80]$,
and the signal to background ratio is
$r=10\% [20\%]$.
These results should motivate a detailed experimental analysis
including Monte Carlo studies~\cite{Bartels:2007ag,Bartelsetal}.

\subsection{Dependence on $\mu$ and $M_2$}

\begin{figure}[t]
\begin{picture}(200,100)
\put(-100,-250){\includegraphics{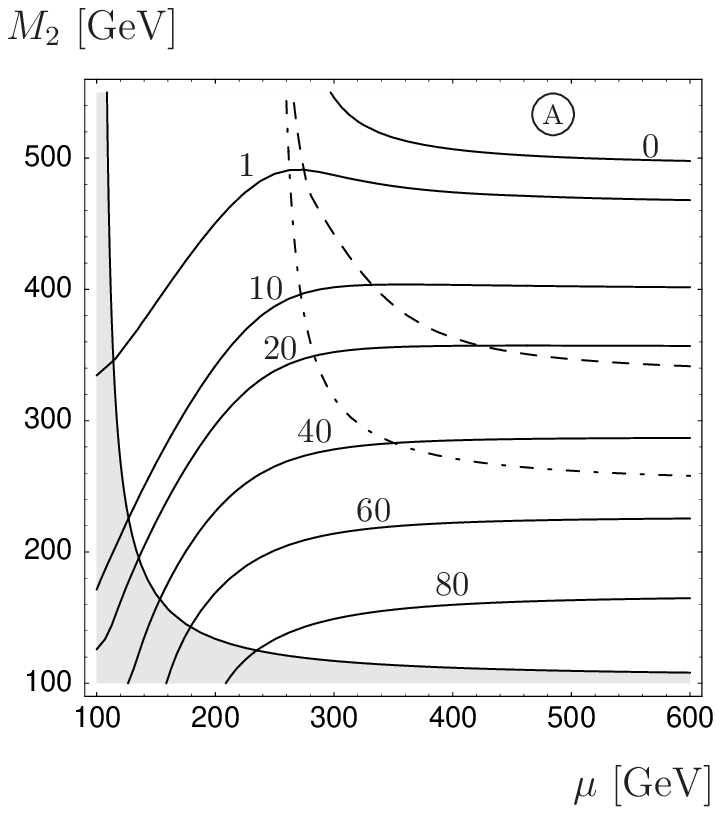}}
\put(40,-250){\includegraphics{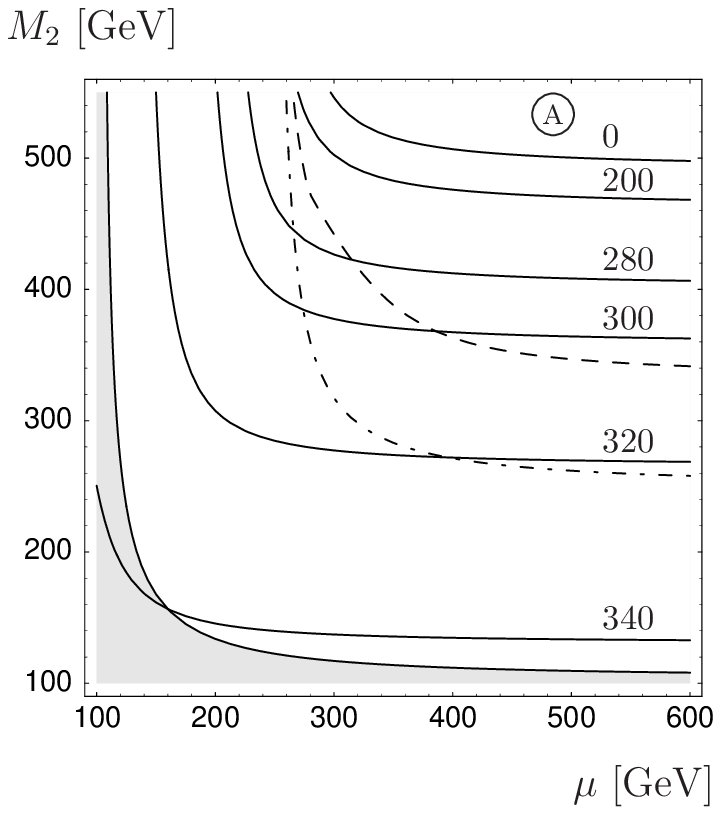}}
\put(180,-250){\includegraphics{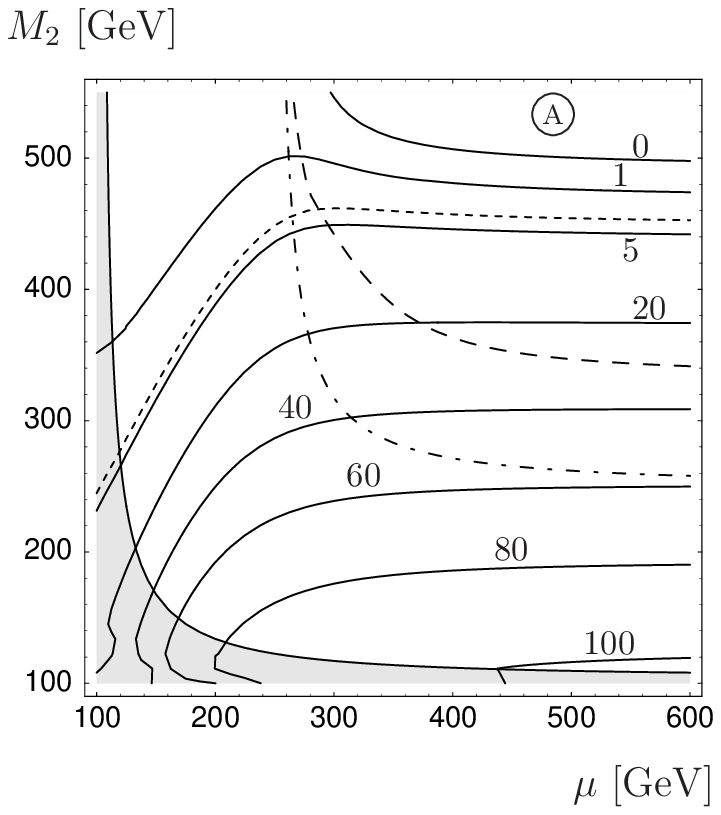}}
\put(20,135){$\sigma(e^+e^-\to\tilde\chi_1^0\tilde\chi_1^0\gamma)$ in $\rm{fb}$}
\put(160,135){$\sigma_{\rm B}(e^+e^-\to\nu\bar\nu\gamma)$ in $\rm{fb}$}
\put(310,135){Significance $S$}
\end{picture}
\caption{\small 
        Dependence of the cross sections and the significance
        on $\mu$, $M_2$ with
        $\sqrt s = 500$~GeV, ${\mathcal L}=500$~fb$^{-1}$,
         $(P_{e^-},P_{e^+})=(0.8,-0.6)$, and
        $\tan \beta=10$, $m_0=100$~GeV~\cite{Dreiner:2006sb}.
        The dashed line
        indicates the kinematical limit $m_{\tilde\chi_1^0}+m_{\tilde\chi_2^0} 
        =\sqrt{s}$, and the dot-dashed line the kinematical limit 
        $2 m_{\tilde\chi_1^\pm} =\sqrt{s}$. Along the dotted line 
        the signal to background ratio is $\sigma/\sigma_{\rm B}= 
        0.01$. The area A is kinematically forbidden by the cut on 
        the photon energy $E_\gamma$, see Eq.~(\ref{cuts}). 
        In the gray area $m_{\tilde\chi_1^\pm}<104$~GeV.
}
\label{fig:mum2}
\end{figure}

In Fig.~\ref{fig:beampol}, we show contour lines 
of the signal $\sigma(e^+e^-\to\tilde\chi^0_1\tilde\chi^0_1\gamma)$
and background cross section $\sigma_{\rm B}(e^+e^-\to\nu\bar\nu\gamma)$
in the $\mu$--$M_2$ plane.
The signal is decreasing for increasing $M_2$,
since the neutralino mass $m_{\tilde\chi^0_1}$ and the 
selectron mass $m_{\tilde e_R}$ are increasing.
For decreasing values of $\mu\lsim300$~GeV, the bino component 
of the neutralino is decreasing, leading to a decreasing signal
cross section $\sigma(e^+e^-\to\tilde\chi^0_1\tilde\chi^0_1\gamma)$.
Note that the background cross section also depends on $\mu$ and $M_2$,
since the kinematical cuts include
the neutralino mass, see Eq.~(\ref{cuts}).
In  Fig.~\ref{fig:beampol}, we also indicate the kinematical
limits of the production of neutralinos $e^+e^-\to\tilde\chi^0_1\tilde\chi^0_2$ 
(dashed line) and charginos $e^+e^-\to\tilde\chi^+_1\tilde\chi^-_1$
(dot-dashed line).
Above these lines, the radiative production of neutralinos is
the only kinematically allowed SUSY state, which can however be observed
with a significance of up to $S=20$.

\subsection{The only state to be observed?}

Consider, as an example, the mSUGRA scenario
$M_0=200$~GeV, $M_{1/2}=415$~GeV, $A_0=-200$~GeV, and 
$\tan\beta=10$. In this scenario, we have 
$M_2=349$~GeV, $\mu=560$~GeV, and
the particle masses are
$m_{\tilde\chi_{1(2)}^0}=180(344)$~GeV, 
$m_{\tilde\chi_1^\pm}=344$~GeV,
$m_{\tilde\tau_1}=253$~GeV,   
$m_{\tilde e_{R(L)}}=261(356)$~GeV.
The neutralinos, charginos
and selectrons are too heavy to be pair produced at
$\sqrt{s} =500$~GeV.
However, neutralinos can still be radiatively 
produced. In  Table~\ref{tab:mSUSGR}, we show the
cross section and the background from
radiative neutrino production for different
sets of beam polarizations with
$P_{e^+}=0,-0.3,-0.6$ and $P_{e^-}=0,0.8,0.9$.
Polarized beams enhance the signal, 
in particular the background is strongly reduced
by a high degree of electron polarization $P_{e^-}=0.9$.
Note that without beam polarization, the signal cannot
be observed.

\begin{table}
\begin{tabular}{|c|cccccc|}
\hline
$(P_{e^+},P_{e^-})$ &$(0|0)$&$(0|0.8)$&$(-0.3|0.8)$&$(0|0.9)$&$(-0.3|0.9)$&$(-0.6|0.8)$\\[1mm]
\hline
$\sigma(e^+e^-\to\tilde\chi^0_1 \tilde\chi^0_1 \gamma)$
&$4.7~{\rm fb}$&$8.2~{\rm fb}$&$11~{\rm fb}$& $8.6~{\rm fb}$&$11.2~{\rm fb}$&$13~{\rm fb}$\\[1mm]
$\sigma_{\rm B}(e^+e^-\to\nu\bar\nu\gamma)$
& $3354~{\rm fb}$ & $689~{\rm fb}$ & $495~{\rm fb}$ & $356~{\rm fb}$&$263~{\rm fb}$& $301~{\rm fb}$\\[1mm] 
$S$ & $1.8$ &$7$ & $11$ & $10$ & $15$& $17$\\[1mm]
$r$ & $0.1\%$ & $1.2\%$ & $2.2\%$ & $2.4\%$ & $4.3\%$& $4.4\%$\\[1mm]
\hline 
\end{tabular}
\caption{Cross sections, significance $S$, and signal to background ratio $r$,
        for different sets of beam polarizations, for
        $\sqrt s = 500$~GeV, ${\mathcal L}=500$~fb$^{-1}$,
        and
        $M_0=200$~GeV, $M_{1/2}=415$~GeV, $A_0=-200$~GeV, 
        $\tan\beta=10$~\cite{Dreiner:2007vm}.
}
\label{tab:mSUSGR}
\end{table}

\subsection{Summary and conclusions}

A pair of radiatively produced neutralinos 
$e^+e^-\to\tilde\chi^0_1 \tilde\chi^0_1 \gamma$
is the lightest
state of SUSY particles to be produced at $e^+e^-$ colliders.
The signal is a single high energetic photon
and missing energy.
The signal could not be observed at LEP due to the 
large background from radiative neutrino production
$e^+e^-\to\nu\bar\nu\gamma$.
At the ILC, however, polarized beams enhance
the signal and simultaneously reduce the background. 
We have shown that the significance for observing
the signal can be as large as $S=100$, and that the signal to background
ratio can be as large as $r=20\%$. These results should motivate
detailed experimental studies, to learn as much as possible 
about Supersymmetry
through the process of radiatively produced neutralinos.


\begin{footnotesize}


\end{footnotesize}


\begin{thebibliography}{99}

\bibitem{url:kittel1} Slides: 
\verb$http://ilcagenda.linearcollider.org/$\\ \verb$contributionDisplay.py?contribId=53&sessionId=69&confId=1296$ and \\
\verb$contributionDisplay.py?contribId=247&sessionId=92&confId=1296$

\bibitem{Haber:1984rc}
  H.~E.~Haber and G.~L.~Kane,
  Phys.\ Rept.\  {\bf 117}, 75 (1985).



\bibitem{ilc}
\verb$http://www.linearcollider.org/wiki/doku.php$

\bibitem{Aguilar-Saavedra:2005pw}
  See, for example, J.~A.~Aguilar-Saavedra {\it et al.},
  Eur.\ Phys.\ J.\  C {\bf 46} (2006) 43
  [arXiv:hep-ph/0511344].



\bibitem{Datta:1994ac}
  A.~Datta, A.~Datta and S.~Raychaudhuri,
  Phys.\ Lett.\ B {\bf 349} (1995) 113
  [arXiv:hep-ph/9411435].

\bibitem{Ambrosanio:1995it}
  S.~Ambrosanio, B.~Mele, G.~Montagna, O.~Nicrosini and F.~Piccinini,
  Nucl.\ Phys.\ B {\bf 478} (1996) 46
  [arXiv:hep-ph/9601292].

\bibitem{Datta:1996ur}
  A.~Datta, A.~Datta and S.~Raychaudhuri,
  Eur.\ Phys.\ J.\ C {\bf 1} (1998) 375
  [arXiv:hep-ph/9605432].

\bibitem{Choi:1999bs}
  S.~Y.~Choi, J.~S.~Shim, H.~S.~Song, J.~Song and C.~Yu,
  Phys.\ Rev.\ D {\bf 60} (1999) 013007
  [arXiv:hep-ph/9901368].

\bibitem{Baer:2001ia}
  H.~Baer and A.~Belyaev,
  [arXiv:hep-ph/0111017].

\bibitem{Datta:2002jh}
  A.~Datta and A.~Datta,
  Phys.\ Lett.\ B {\bf 578} (2004) 165
  [arXiv:hep-ph/0210218].

\bibitem{Dreiner:2006sb}
  H.~K.~Dreiner, O.~Kittel and U.~Langenfeld,
  Phys.\ Rev.\  D {\bf 74} (2006) 115010
  [arXiv:hep-ph/0610020].

\bibitem{Dreiner:2007vm}
  H.~K.~Dreiner, O.~Kittel and U.~Langenfeld,
  arXiv:hep-ph/0703009.

\bibitem{Wilson:2001me}
  G.~W.~Wilson,
   LC-PHSM-2001-010.

\bibitem{Birkedal:2004xn}
  A.~Birkedal, K.~Matchev and M.~Perelstein,
  Phys.\ Rev.\ D {\bf 70} (2004) 077701
  [arXiv:hep-ph/0403004].

\bibitem{Bartels:2007ag}
  C.~Bartels, Diploma Thesis, 
  \emph{Model-independent WIMP searches at the ILC},
DESY 2007. 

\bibitem{Moortgat-Pick:2005cw}
  G.~A.~Moortgat-Pick {\it et al.},
  arXiv:hep-ph/0507011.


\bibitem{Allanach:2002nj}
  B.~C.~Allanach {\it et al.},
in {\it Proc. of the APS/DPF/DPB Summer Study on the Future of Particle Physics (Snowmass 2001) } ed. N.~Graf,
  Eur.\ Phys.\ J.\ C {\bf 25} (2002) 113
  [eConf {\bf C010630} (2001) P125]
  [arXiv:hep-ph/0202233].

\bibitem{Bartelsetal}
  C.~Bartels et al., in preparation.


\end{thebibliography}
\end{document}